**TITLE**

On Borrowed Time: The Past, Present, and Future of Virginia's Barrier Islands under Differing Sea-Level Rise Scenarios.

**AUTHORS**

Stuart E Hamilton[+,^], John Talbot[+], Carl Flint[+], Adam Phipps-Dickerson[+], Tyler Wilson[+], Mychael Smith[+].

[+] Dept. of Geography & Geoscience, Henson Science Hall, Salisbury University, 1101 Camden Av., Salisbury MD, 21801, USA

[^] Corresponding author, sehamilton@salisbury.edu

**ABSTRACT**

      Virginia's barrier islands constitute one of the most undeveloped shorelines of the eastern US. Aside from a few islands in the north, the islands are uninhabited and managed for conservation. These islands play important environmental, cultural, and economic roles along Virginia's Eastern Shore. Climate change driven sea-level rise is already having a major impact on these islands and threatens their existence. We utilize transect analysis across each of the barrier islands to depict the shoreline change trends annually from 1850 to 2010. We then utilize time series forecasting and panel modeling to estimate future island shorelines up to and including a best estimate 2099 CE shoreline. Results indicate that across almost all the islands, the shoreline retreat rate has been increasing over time. Additionally, we find that year 2100 CE sea-level rise scenarios are likely to accelerate the shoreline retreat occurring on these islands and may erase many of them all together. We find that the northern islands of Wallops and North Assateague will remain generally stable whereas many of the remaining islands will likely experience rapid shoreline retreat under future sea-level rise scenarios.

**KEYWORDS**

      Virginia's Barrier Islands, Sea-Level Rise, Shoreline Change, Climate Change

**INTRODUCTION**

The barrier islands of Virginia play an important ecological role along Virginia's Eastern Shore (VES) in addition to being of regional economic and cultural importance. For example, the barrier islands of VES provide valuable wildlife habitat (Ayers, 2005; Rood, 2012; U.S. Fish & Wildlife Service, 2015) support regionally important commercially fisheries (Kirkley, 1997; Richards & Castagna, 1970), provide economically important tourism & recreational opportunities (Rood, 2012), and likely provide protection to the mainland from storm events. From an economic perspective, the populations of VES depend heavily on the existence of these barrier islands. The landward lagoon ecosystems, that enhances fisheries for a multitude of finfish and shellfish species (Kirkley, 1997) would not exist, or be heavily altered, without barrier islands. For example, the coastal lagoons of VES play an important role in the regional commercial clam, oyster, finfish, scallop, conch, and blue crab fisheries (Kirkley, 1997).

Both Accomack and Northampton counties that constitute the VES, rank among the highest in poverty and unemployment levels within Virginia as well as having incomes substantially below the state average (U.S. Department of Agriculture: Economic Research Service, 2015). The VES barrier islands provide numerous economic good and services to these counties, that if lost may increase levels of hardship in this peripheral region. For example, the sheltered lagoons provide mooring locations for private and commercial fishing fleets from Oyster in the south to Chincoteague in the north. These lagoons additionally contribute directly to employment and economic activity through the commercial seafood capture, processing, and distribution sectors as well as supporting numerous aquaculture ventures. For example, both VES counties have seafood processing and distribution industries and Accomack County recently built the Robert S. Bloxom Shore Agricultural Complex, a 35,000 ft$^2$ facility designed to support the local seafood, clam, fish, and oyster markets (Accomack County Economic Development Authority, 2015).

In addition to the market sectors usually associated with fisheries, many specialized jobs also exist on VES related to the fisheries industry. For example, specialized insurance agents are required to deal with issues of maritime insurance (Kirkley, 1997), specialized mechanics are needed to deal with pumps within the aquaculture industry, and numerous fisheries research activities are conducted in the area (Rood, 2012). From a tourism perspective, the communities of Wachapreague, Assateague, and Chincoteague that reside on the interior of these lagoons are all recognized as regionally significant tourist centers. An additional tourist component of the VES barrier islands lagoons system is the opportunity they provide for nearshore-fishing as part of the large recreational fishing market with small craft boat launches stretching the entire length of the lagoon system (Ayers, 2005).

The economic importance of VES barrier island environments and their associated inland waterways are not limited to fisheries, tourism, and recreation. One less documented economic activity, likely driven by the a combination of the remoteness and the pristine nature of the islands, is the development of a cluster of research & teaching facilities (Rood, 2012) in the region. For example, VES Wallops Island is home to NASA's largest satellite launch facility and is the only rocket launch range currently operated by NASA with a history of 16,000 rocket launches (Young, 2015). Further south, Virginia Institute for Marine Science Eastern Shore Laboratory (VIMS) is located on one landside of one the lagoons in Wachapreague, VA. In addition to NASA and the VIMS campus, Salisbury University, University of Virginia, University of Maryland Eastern Shore, University of Maryland, Virginia Tech, and the Barrier Islands Center all have education or research programs in the area that are either directly or indirectly related to the barrier islands and their associated lagoons (Rood, 2012).

The ecological, cultural, and economic past, present, and future of the barrier islands of VES are under threat from climate change induced sea-level rise (SLR) (Crichton, 2014; Strange, 2008). The entire VES barrier island system is reported as being vulnerable to even modest levels of SLR, with almost total inundation of the islands possible in a 1 m SLR scenario (Crichton, 2014). With the exception of a few northern islands that take engineering-based mitigation action (King, Ward, Williams, & Hudgins, 2010), the entire SLR barrier islands are managed for conservation (Figure 1) and engineered sea defenses are unlikely to be constructed. SLR is not only a predicted future problem for the barrier islands of VES; they are already experiencing shoreline retreat (SLRT) and erosion due to SLR already occurring. For example, the Atlantic Ocean facing Wreck Island shoreline is reported to have retreated 300 ft. in recent years and sea-level rise is reported to be responsible for as much as 40 ft. of island SLRT in a single year (Rood, 2012). Indeed, many of the dramatic photographs of homes in the ocean that accompany SLR news stories in the popular press are actually abandoned properties along the barrier islands of VES (see Ayers 2005 for a sample).

The academic community appears to be reaching consensus on future SLR and a predicted 2100 global sea level approximately 1 m greater than present (Table 1). Some research outliers still occur such as Houston and Dean (2011) who find that 21st Century tide data measures taken in isolation depict a trend one to two orders of magnitude less than required to meet many of the year 2100 SLR predictions represented in Table 1. Despite some findings to the contrary, an overwhelming number of post-2008 primary research publications support the year 2100 1 m SLR scenario represented in Table 1. Even the conservative estimates of climate change induced SLR depicted in Table 1 predict that SLR will substantially outpace the current global rate of 3.2 mm/yr$^{-1}$ of SLR since the early 1990s (Church et al., 2013; Merrifield, Merrifield, & Mitchum, 2009). As SLR is the primary driver of barrier island erosion and SLRT (Bruun, 1962; Leatherman, 1988; Leatherman, Zhang, & Douglas, 2000), it appears sensible to assume increased levels of both erosion and SLRT for the barrier islands of VES under all future SLR scenarios and across all models (Table 1).

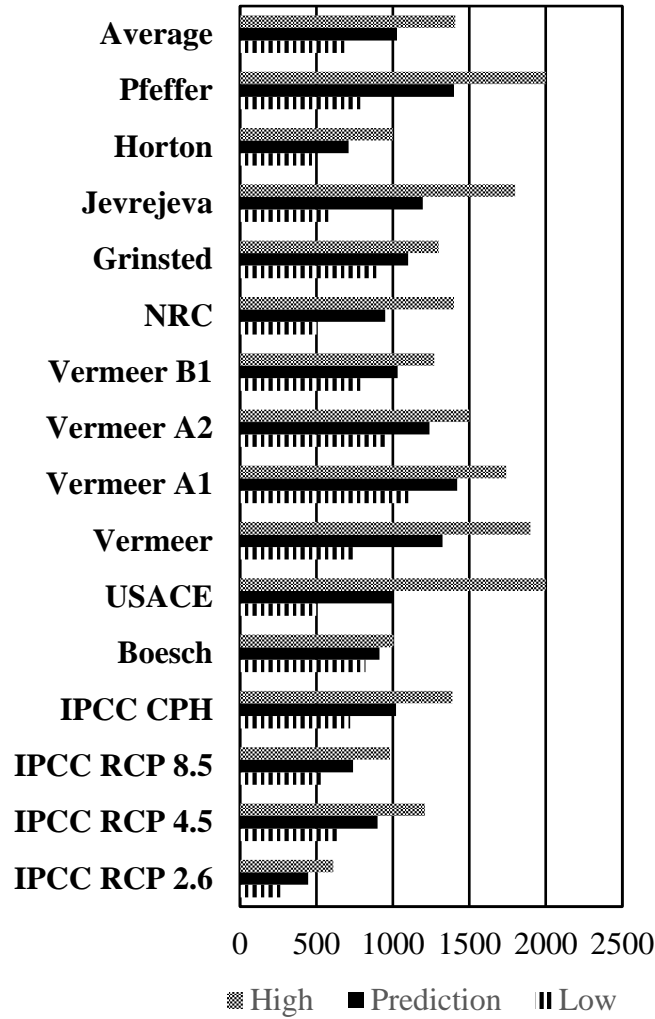

**Table 1.** Year 2100 Predicted Sea Levels and Uncertainty (Boesch et al., 2013; Church et al., 2013; Grinsted, Moore, & Jevrejeva, 2009; Horton et al., 2008; Jevrejeva, Moore, & Grinsted, 2010; National Resource Council, 2012; Pfeffer, Harper, & O'Neel, 2008; U.S. Army Corps of Engineers, 2011; Vermeer & Rahmstorf, 2009). All units have been adjusted from their reported values to mm for consistency. The solid black bars represent reported SLR by the author or if not reported the mid-value of their reported range. The high value indicates the reported range on the high end, and the low value indicates the reported range on the low end. The authors reported their ranges and error in different methods so a rescaling was necessary for consistency. All predictions state that the year of prediction is 2100, 2099, the end of the twenty-first century or similar. The start dates show variation but generally indicate 2000, present, or similar. The sample is a sample of convenience of the academic literature available to a US-based researcher. These estimates cannot be assumed independent of each other and many are likely related or based on the same core dataset. The IPCC 5[th] assessment SLR calculations are assumed to supersede the values reported in the four earlier assessments that are therefore omitted.

If the generalized relationship of between 1 part SLR to 100 - 150 parts SLRT (Bruun, 1962; Leatherman et al., 2000; Watson, Zinyowera, & Moss, 1998) continues into the future, and Table 1 is correct, then the barrier islands of VES and the associated lagoon systems may already predestined to disappear. Indeed, the IPCC findings indicate that even if global carbon emissions are cut to zero overnight then we are already committed to between 400 and 800 mm of average global SLR (Church et al., 2013) which would likely erode or submerge many of the islands. For example, in a scenario of future SLR matching twentieth-century SLR (Table 1, Average Low) it is virtually certain that Assateague Island, which is likely already at geomorphic threshold, will be severely degraded as to no longer function as a barrier island (Gutierrez, Williams, & Thieler, 2007). By the moderate consensus SLR (Table 1, Average) it is as likely as not that a geomorphic threshold would be reached across numerous islands and lead to island disintegration (Gutierrez et al., 2007). By the high-level consensus (Table 1, Average High) it is virtually certain that threshold behavior will be seen across the island chain (Gutierrez et al., 2007).

The barrier islands of VES may be in an even more precarious position that other barrier islands globally as SLR along VES has historically exceeded the global SLR averages. For example, relative sea-level rise in Virginia during the twentieth century is recorded as 4.4 mm/y[-1] (C Zervas, Gill, & Sweet, 2013), while the consensus average global rate of over the same period is 1.7 mm/yr[-1] ±0.5 (Bindoff et al., 2007). Indeed, the barrier islands of VES may not only be experiencing higher rates of SLR than other regions globally, but also higher rates than other states along the US eastern seaboard (Boon, 2012; Ezer & Corlett, 2012; Grinsted et al., 2009; Sallenger Jr, Doran, & Howd, 2012; U.S. Army Corps of Engineers, 2011). This is likely in part due to vertical land movement (VLM). Within the VES barrier island region, VLM is reported as between -1.3 mm/yr$^{-1}$ to -2.7 mm/yr$^{-1}$ annually (Boon, Brubaker, & Forrest, 2010; Grinsted et al., 2009; Holdahl & Morrison, 1974; C Zervas et al., 2013) across varying long-term periods. The literature is unclear if VLM contributes to barrier island SLRT in addition to mainland SLRT due to the geologic construction of these mostly sand features. One additional factor that may support accelerated SLRT of the barrier islands of VES when compared to many other areas globally is that the islands reside in a tropical storm region and barrier islands are known to retreat, often substantially, during tropical storm events (Houser & Hamilton, 2009; Houser, Hamilton, Meyer-Arendt, & Oravetz, 2007; Houser, Hapke, & Hamilton, 2008). Future tropical storms, under warmer climate scenarios, are predicted to increase in magnitude (Knutson et al., 2010) if not frequency (Knutson et al., 2010).

As opposed to merely accounting for SLR by increasing sea level by a predicted amount and subtracting this from the current terrain surface, we approach the question of SLR by examining SLRT under differing SLR scenarios. In this paper, we measure the location of the past and current shorelines of the barrier islands of VES. We then use these time-series data to make future predictions of shoreline location without the use of covariates under the assumption that the relationship between past SLRT and SLR will be the same as the future relationship. We then use the high-resolution historic shorelines created for the time-series analysis to extract the localized relationships between SLRT and SLR along the barrier islands. We then use these historic relationships to predict SLRT scenarios through to 2100 using the time-series SLR data, under a range of differing climate-driven scenarios as reported in the academic literature.

We use both longitudinal and panel methods to generate our predictions within a GIS environment. The analysis utilizes 1463 high-resolution transects, between 10 and 14 historic shorelines per island, and approximately 18,000 observation points across all twenty-three islands. The findings are reported at the island and regional levels. This analysis is likely the highest resolution localized examination of barrier-island response to future SLR yet undertaken. The result produced are only applicable to this region but the methods presented provider wider value to those studying issues of barrier island morphology under future SLR scenarios. The final product is a best-fit map of the future shorelines of the VES barrier islands in 2100.

**STUDY AREA**

The barrier islands of VES exist from the mouth of the Chesapeake Bay in the south to the border with Maryland in the north (Figure 1). They consist of approximately 178 km of shoreline, approximately twenty-three islands, and have a combined area of approximately 68,500 km$^2$ (Barrier Islands Center, 2015). The major named islands[1] as you move from north to south are Assateague, Wallops, Assawoman, Metompkin, Cedar, Parramore, Hog, Cobb, Wreck, Ship Shoal, Myrtle, and Smith. With the exception of the northern islands, which contain the NASA Wallops Island launch facility and the city of Wachapreague, the majority of the islands are uninhabited and managed for conservation with a patchwork of differing environmentally conscious owners managing the islands under various levels of protected status (Ayers, 2005) (Figure 1). Indeed, the Nature Conservancy list the barrier lists of VES as one of their long-term flagship programs having receiving over $100 million to protect and restore the barrier islands and their lagoons (Crichton, 2014).

---

[1] For the purposes of this paper, we do not include Fisherman Island as barrier island as it a circular island that exists in the mouth of the Chesapeake Bay and not offshore in the Atlantic Ocean.

**Figure 1.** Study Area and Barrier Island Ownership and Conservation Regimes.

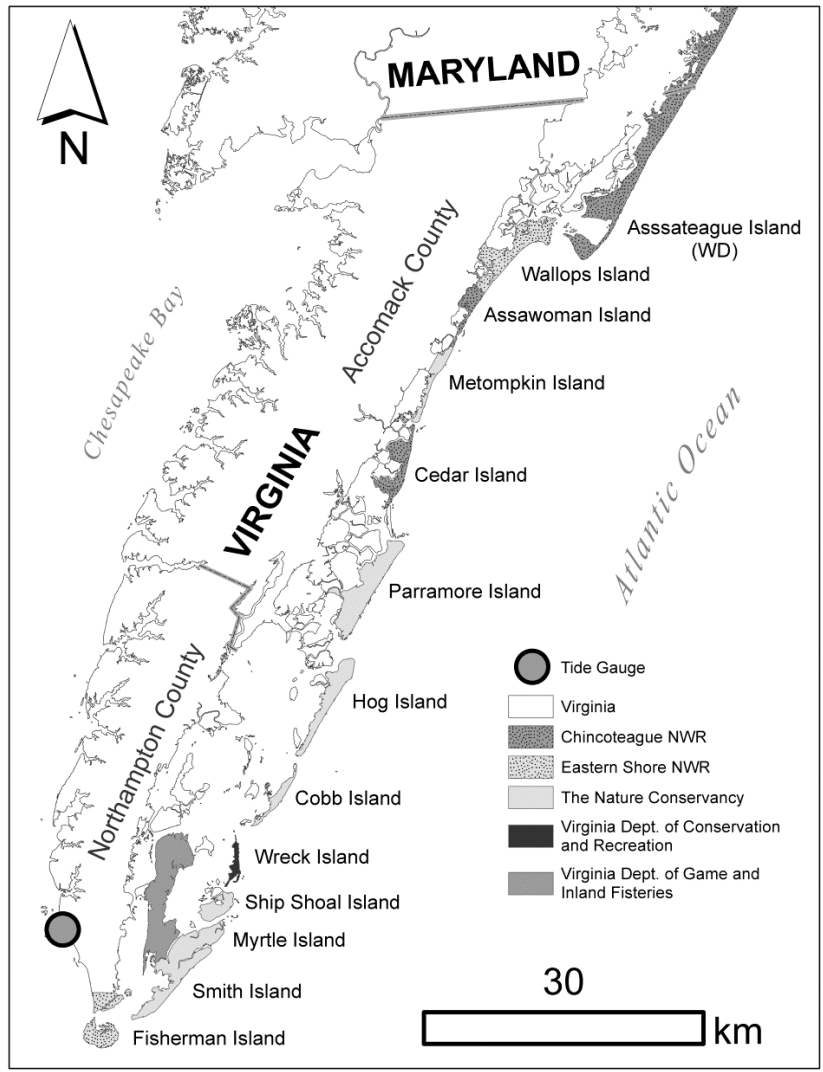

WD= wave-dominant islands, the renaming islands are mixed-energy barrier islands (Gutierrez et al., 2007; Hobbs III, Krantz, & Wikel, 2015; Leatherman, 1988). Cedar Island is part owned by The Nature Conservancy and partly in private ownership. The tide gauge used in is this analysis is across the Chesapeake Bay at Sewell's Point due to a longer time record.

The VES barrier island constitute one of the most undeveloped shorelines of the eastern US, are the longest global stretch of undeveloped barrier islands in the temperate zone, and are recognized as a UN designated Biosphere Reserve (Barrier Islands Center, 2015). The VES barrier islands additionally play an important role in international bird migration and habitat with, "the barrier island ecosystem sheltering more than 250 species of raptors, shorebirds, and songbirds and is (sic) one of the most important migratory bird stopover habitats on Earth" (Rood, 2012). Indeed, many millions of birds utilize these islands annually, with the lagoon ecosystems providing important habitat as part of the Atlantic Flyway migration system (Virginia Department of Conservation and Recreation: Virginia Natural Heritage Program, 2015).

**GEOLOGIC FRAMEWORK**

The geologic and geomorphological frameworks of the eastern shore and their offshore islands are well documented in the academic literature. Hobbs III et al. (2015), provide a thorough geomorphic review of VES barrier island system and discuss most of the barrier islands and inlets individually. Mixon (1985), provides a geologic framework of the entire Delmarva Peninsula describing the relatively modern formation of VES. Demarest and Leatherman (1985), provide a geologic analysis of each of the coastal systems identified on Delmarva Peninsula including the VES barrier island system. Colman, Halka, Hobbs, Mixon, and Foster (1990), provide a discussion of the paleo-channels that now form the major navigation channels between the VES barrier islands. At the local level, Newman and Munsart (1968), provide a geologic framework for the Wachapreague lagoon and Shideler, Ludwick, Oertel, and Finkelstein (1984), analyze the quaternary stratigraphic evolution of the southern portion of the study area. This paper is limited to a discussion of the key portions of the geologic and geomorphic frameworks relevant to the study.

In general, barrier islands are classified into wave-dominated barrier islands (WDBI), —

sometimes referred to as mesotidal—and mixed-energy barrier islands (MEBI)—sometimes

referred to as macro tidal— (Hayes, 1979; Leatherman, 1988). WDBI are generally long and thin

stretches of sand in front of lagoons with wave energy generally dominating tidal energy and are

susceptible to overwash (Hayes, 1979). MEBI are generally less linear and have a greater

distance from their exterior shoreline (Hayes, 1979). The VES barrier islands are a relatively

geologically young formations likely occurring within the last 5,000 – 10,000 years via processes

of longshore drift and accretion (Schwartz 2005), as the Delmarva peninsula slowly expanded

southwards creating the Chesapeake Bay in the process (Mixon, 1985). Since their formation the

barrier islands locations have fluctuated due to sediment supply, hydrodynamic conditions, as

well as atmospheric phenomena (Hayes, 1979). The extreme northern barrier islands of VES are

WDBI and the remaining islands as MEBI (Gutierrez et al., 2007; Hobbs III et al., 2015;

Leatherman, 1988) (Figure 1).

**MATERIALS AND METHODS**

As recently as 2007, a general lack of consensus existed among coastal scientists on the use of appropriate methodologies to monitor the coastline implications of SLR (Gutierrez et al.,

2007). Many regional SLR analyses conducted along the Atlantic Coast of the US utilize DEMs or DTMs as a current zero baseline and then utilize various techniques that essentially add differing SLR scenarios to these base levels and utilize these baseline plus SLR additions to show future inundation areas (e.g. Gesch, 2009; Strauss, Ziemlinski, Weiss, & Overpeck, 2012;

Wu, Najjar, & Siewert, 2009). Such approaches have importance in alerting regions to their potential inundation under differing SLR scenarios but they do not account for the complex interaction between SLR and SLRT that dominates the morphology of barrier islands.

To account for historic SLRT and differing future potential SLR scenarios we use two different methods in this paper to predict SLRT along the barrier islands of VES. The first approach predicts future changes in their shoreline location using an empirical model based on past observations of SLRT in this region during the last 160 years. Shoreline change for each of the barrier islands of VES are calculated over three different historical periods and these results are then extrapolated into differing rates used to make future predictions. These periods are the long-term erosion rate based on shoreline change data from 1850 to 2010, the medium-term erosion rate based on shoreline change data from 1925 to present, and the current shoreline erosion rate based on shoreline change data from 1970 to present. These rates are then used to model future shoreline location under a business-as-usual scenario. The second approach accounts for the future expected changes in SLR that deviate substantially from earlier LR scenarios.

Barrier island shoreline migration can be loosely defined as being a function of SLR, sediment supply (SS), wave energy (WE), and human intervention (HI) (Equation 1).

Equation (1)

$$SM = f(SLR\ SS\ WE\ HI)$$

SM = shoreline migration. SLR = sea-level rise. SS = sediment supply, often as a function of longshore transport. WE = wave energy, most often as a function of storm events. HI = human intervention, such as the beach nourishment, the building of groins, jetties, and hardened shoreline defense's (Shao, Young, Porter, & Hayden, 1998).

We assume that Virginia's Barrier Islands HI remains constant. That is, the region around Wallops Island and Assateague Island will continue to have their shoreline defenses maintained and no additional shoreline defenses will be built along the remaining islands that are managed for conservation. This matches the predicted future shoreline defense scenarios for this region as defined by the EPA report examining the likelihood of shore protection along the Atlantic Coast of the US (Titus et al., 2010). Additionally, we assume that SS remains constant into the future with a similar longshore transport system in the future as to what exists in the past. We identify SS as a future research need, as little appears to have been done accounting for future SS under differing SLR scenarios. We then vary both SLR based on future SLR scenarios and WE based on future storm frequency predictions.

To monitor past shoreline change and predict future shoreline change we divided the barrier islands of VES into thirteen logical individual islands. We then created a linear survey line for each of the thirteen islands parallel to the shoreline. This survey line is located between each island and the mainland and is used as a baseline to measure past, present, and future shoreline erosion or accretion. We then subdivided each of the 13 islands into 100 m segments and utilized these 100 m intervals as the locations for our transect analysis, this resulted in 1463 transects across the study area. The 100 m spaced transects start at the landward survey line and extend beyond the most distant historic shoreline in a manner that is approximately perpendicular to the majority of each of the 13 shorelines. Once the shorelines were digitized or downloaded these transects served as the basis for the regression and endpoint analysis (EPR) on the shorelines that followed.

The shoreline start dates for each island is on-or-around 1850 (Table 2) and the universal end date is 2010. Regression analysis involved extracting each shoreline for each period (Table 2) at each transect and using a linear regression to make future shoreline predictions. EPR analysis is merely the extraction of the extreme landward shoreline and the extreme seaward shoreline and calculating the rate of change between these endpoints and making future predictions based on this rate of change. The shoreline regressions and the EPR return not only annual rates of change across each transect but the regression approach also returns standard error margins and confidence intervals (Thieler, Himmelstoss, Zichichi, & Ergul, 2009) and hence are preferred in this analysis. The regression and EPR are conducted for three temporal domains; from 1850 to 2010 for long-term shoreline change rate, from 1925 to 2010 for medium-term shoreline change rate, and 1970 to 2010 for the current rate. The data were also collapsed by space as well as time, with each transect weighted equally for each island. Future shoreline location predictions were made based on the long-term trend, medium-term trend, and current trend of historical shoreline change. That is, the three temporal domains are assumed to continue into the future and are modelled through to 2100.

| Island Name | Shorelines Digitized | Transect Count |
|---|---|---|
| Assawoman | 1851, 1855, 1910, 1933, 1942, 1943, 1962, 1980, 1997, 2002, 2006, 2010 | |
| Cedar | 1852, 1871, 1910, 1942, 1962, 1980, 1997, 2002, 2006, 2010 | |
| Assateague North | 1850, 1859, 1908, 1933, 1942, 1962, 1980, 1989, 1997, 2000, 2006, 2010 | |
| Assateague South | 1908, 1915, 1933, 1942, 1943, 1962, 1979, 1980, 1997, 2000, 2006, 2010 | |
| Cobb | 1853, 1870, 1911, 1934, 1942, 1962, 1980, 1997, 2002, 2006, 2010 | |
| Hog | 1852, 1853, 1871, 1911, 1919, 1934, 1942, 1962, 1967, 1980, 1997, 2002, 2006, 2010 | |
| Metompkin | 1851, 1852, 1855, 1910, 1933, 1942, 1962, 1980, 1997, 2002, 2006, 2010 | |
| Myrtle | 1852, 1888, 1911, 1921, 1942, 1962, 1980, 1997, 2002, 2006, 2010 | |
| Parramore | 1852, 1871, 1910, 1911, 1942, 1962, 1980, 1997, 2002, 2006, 2010 | |
| Ship Shoal | 1852, 1888, 1911, 1942, 1962, 1980, 1997, 2002, 2006, 2010 | |
| Smith | 1852, 1888, 1905, 1911, 1921, 1942, 1943, 1953, 1962, 1980, 1997, 2002, 2006, 2010 | |
| Wallops | 1851, 1855, 1887, 1910, 1915, 1933, 1943, 1962, 1979, 1980, 1997, 2002, 2006, 2010 | |
| Wreck | 1852, 1888, 1911, 1942, 1962, 1980, 1997, 2002, 2006, 2010 | |
| Total | | |

**Table 2.** Year of shoreline digitization and transect count as of 2010 for each island.

To predict year 2100 SLRT scenarios from past data we aggregated the SLRT data into island / decadal averages and then we then calculated the SLR rate based on data from 1927 to present from the Sewell's Point tide gauge and aggregated these SLR measures to island / decadal averages (Figure 2). These data reflect an almost linear SLR since the 1920s with a slight increasing trend as the data moves towards 2010. These data provide the historic SLR data information that corresponds to the decadal shoreline change information. We additionally project these data using a simple linear OLS forecast of all historic data points (Figure 2, MSL) and the three average scenarios outlined in Table 1 reflecting the higher average 2100 SLR level

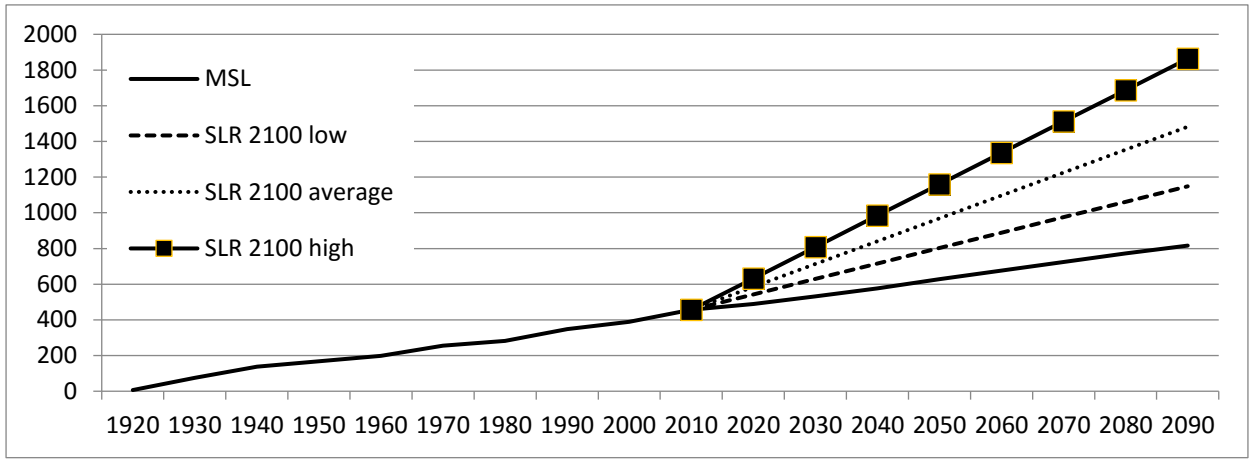

**Figure 2.** MSL in mm with 1927 set to zero. Data pre-2010 is measured data and data post-2010 is projected. The solid black line represents SLR continuing on its post-1927 trajectory through until 2100, this is the business-as-usual scenario. The dashed line represents the low average SLR prediction from Table 1. The dotted line represents the average SLR prediction from Table 1. The line interrupted with black squares represents the high average SLR prediction from Table 1.

Using simple linear regression at the island / decadal level we find that the majority of the barrier islands erosion depicted in Table 1 can be explained by SLR and storm events. This finding reinforcing the current barrier island shoreline change and erosion paradigms (Bruun, 1962; Houser & Hamilton, 2009; Houser et al., 2008; Leatherman, 1988; Leatherman et al., 2000). Once the historic relationship between the islands shoreline change and SLR combined with WE was ascertained we were able to manipulate SLR and make year 2100 shoreline location predictions based on differing SLR scenarios as depicted in Figure 2. That is, SLR was varied and the relationship between SLR and SLRT was applied into the future based on how shoreline change had responded to SLR in earlier periods. This allowed for creation of a 2100 shoreline along the entire coastline based on data at the island level under differing SLR scenarios.

All past and future shoreline change estimates are provided with a measure of uncertainty. We utilize the standard error of the slope methodology (Thieler et al., 2009), reported as the 90 percent confidence level. The standard error of slope is constructed using a best-fit regression through each of the individual transect points across each of the 1463 transects, the slope of the line is the rate of change in $m/yr^{-1}$. The confidence interval is then calculated by multiplying the standard deviation —standard error of the sample points along the slope—in a classic two-tailed Z-test (Zar, 1999).

**SOFTWARE AND DATA SOURCES**

Hourly tide data from 1927 to present was obtained from the NOAA Tides and Currents product for the NOAA Sewell's Point tide station (8638610). SLR predictions are taken from the various sources listed Table 1. Historic shoreline data from 1850 to 2000 was retrieved from the USGS National Assessment of Shoreline Change (Miller, Morton, & Sallenger, 2005). Shoreline data for the 2000s were manually digitized utilizing high-resolution imagery obtained from the USGS imagery program. Lidar data was obtained from FEMA to create a 2010 shoreline. The software used was ArcGIS 10.3.1, DSAS (Thieler et al., 2009), Python 2.7, QGIS, GDAL, and Microsoft Excel. All geographic data was processed from various projections into EPSG: 32618. The vertical definition of zero is the MLLW measure of zero at NOAA tide gauge ID 8638610, which corresponds to 356.62 mm below NGVD29 zero. All GIS, statistical, and tabular data required for replication; including shorelines, transects, regressions, code, and error analyses are available without restriction from the Harvard Dataverse at https://dataverse.harvard.edu/dataverse/BIVES and researchers are encouraged to utilize these data or replicate the findings of this research using these data.

**RESULTS**

Across the thirteen major islands and three temporal domains, three major historic shoreline change trends are identified. The first is stability, the most northern islands of Assateague North and Wallops show little or no SLRT across any of the three temporal domains, they are essentially stable (Figure 3). The second dominant pattern is one of increasing levels of SLRT as you move towards present. The islands of Myrtle, Cobb, Parramore, Cedar, and Assawoman show a general pattern of high levels of SLRT that are increasing in magnitude towards present (Figure 3). The final dominant pattern is one of SLRT but with no consistent temporal pattern. This pattern applies to Smith, Ship Shoal, Wreck, Hog, and Metompkin Islands (Figure 3).

**Figure 3.** SLRT by temporal domain and island. A value of zero means no SLRT occurred or limited accretion. The units are m/y$^{-1}$. Units are meters per year. The CI noted are 90 percent. Wreck's error bar extends to 27 m.

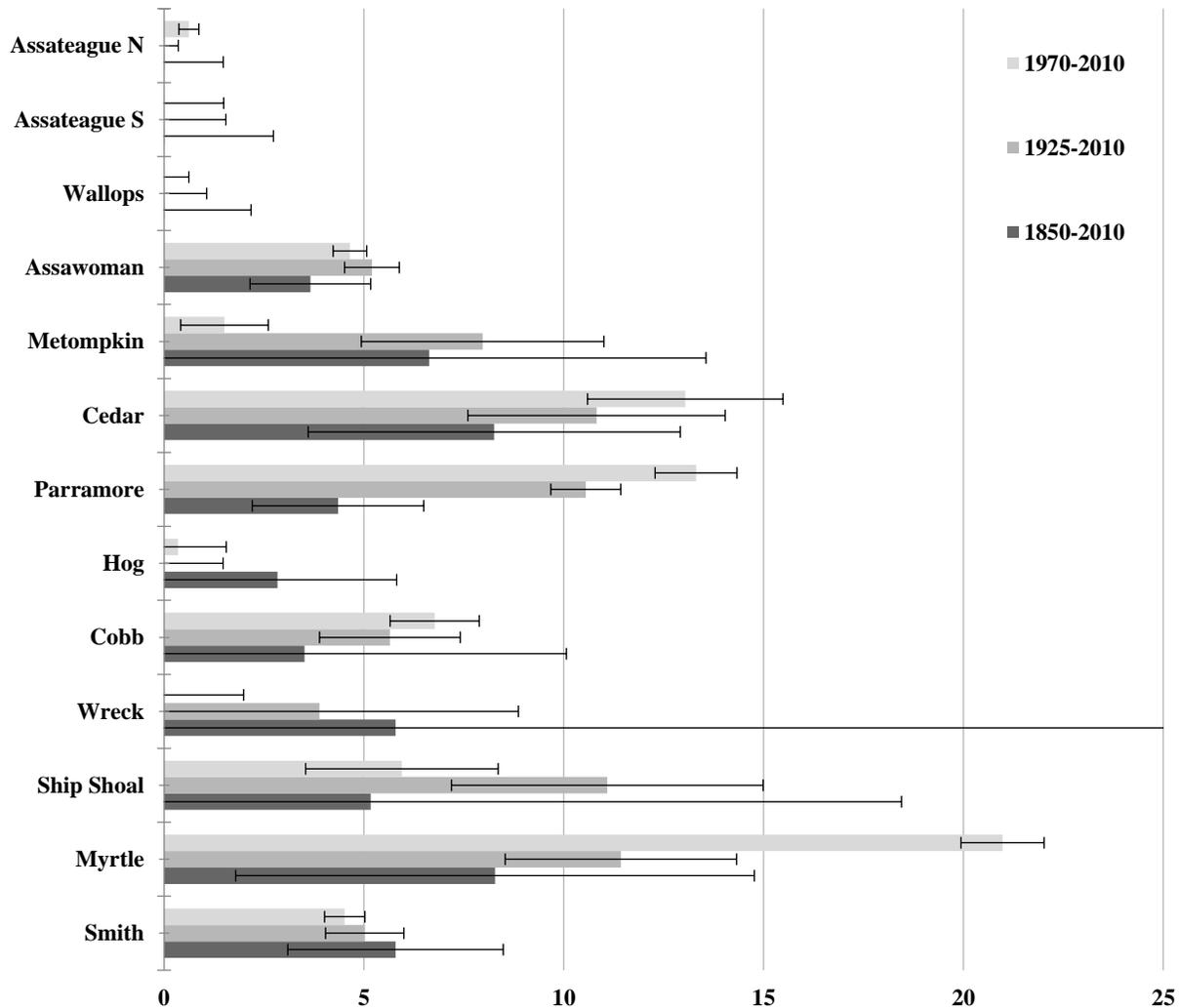

Figure 3 reflects that SLRT is generally consistent across all non-mitigated islands over all three temporal domains. Error bars are depicted at the 90 percent confidence interval. In addition to being consistent, SLRT appears to be increasing as you move towards present. For example, the period from 1850 – 2010 averages 5.5 m/yr$^{-1}$ of SLRT per island but this increases to 7.1 m/yr$^{-1}$ for both the later domains. In addition to average SLRT increasing as the dates move closer to present, the number of islands with extreme SLRT also increases. For example, the long-term epoch of 160 years has no individual island SLRT greater than 8.5 m/yr$^{-1}$, yet the mid-term and current epochs have four islands with an average SLRT greater than 10 m/yr$^{-1}$, in addition to the first appearance of extreme values averaging beyond 12 m/yr$^{-1}$ (Figure 3). The appearance of extreme values adds evidence to the potential exponential and non-linear relationship between SLR and SLRT (Bruun, 1962; Leatherman et al., 2000; Watson et al., 1998). If we isolate 1850 to 1925, 1925 to 1970, and 1970 to 2010, then the SLRT values average 5.3 m/yr$^{-1}$, 6.8 m/yr$^{-1}$, and 7.1 m/yr$^{-1}$ respectively. This increases the level of certainty the SLRT is increasing while transiting through the three temporal domains from 1850 to present.

Extending the three historic temporal domains (Figure 3) from 1850, 1925, and 1970 to 2100 results in a substantial increase in future SLRT in all scenarios as the more recent temporal domains (Figure 3) exhibit more SLRT that the 1850 – 2010 domain (Figure 3). That is, the pattern of SLRT is not only high across all domains but also increasing in more recent times. This increasing of SLRT is likely due to the well-documented recent relative SLR increases in this region (Boon, 2012; Ezer & Corlett, 2012; Grinsted et al., 2009; Sallenger Jr et al., 2012; U.S. Army Corps of Engineers, 2011; Chris Zervas, 2001; C Zervas et al., 2013).

The year 2100 map of barrier island of VES will look unrecognizable from today if the 1970 – 2010 SLRT (Figure 3) pattern continues through to 2100. For example, Myrtle Island is approximately 10 km from the mainland and 8 km from Magothy Bay at present. If SLRT continues through to 2100 CE, then this shoreline is predicted to retreat to the mainland and the island will cease to exist (Figure 4). It is not only the islands with high levels of historic SLRT that are likely to disappear but also many others with more average levels of historic SLRT. For example, parts of Cedar Island are only <3 km away from the mainland but the 2100 CE predicted shoreline of Cedar Island will retreat the majority of this distance causing at least large portions of the island to disappear. In addition to Cedar Island and Myrtle Island, Metompkin Island and Assawoman Island appear to be at risk of full retreat to the mainland by 2100. These extreme island projections are based on the 1970 - 2010 historic shoreline retreat data, and reflect a business as usual scenario and do not account for future more rapid increases in SLR induced SLRT that would likely occur under any of the scenarios depicted in Table 1.

**Figure 4.** Total shoreline retreat in kilometers by 2100. The trend is based is standard OLS regression using the 1970 – 2010 temporal domain extended through to 2100. Wallops Island SLRT is adjusted to zero but shows slight accretion.

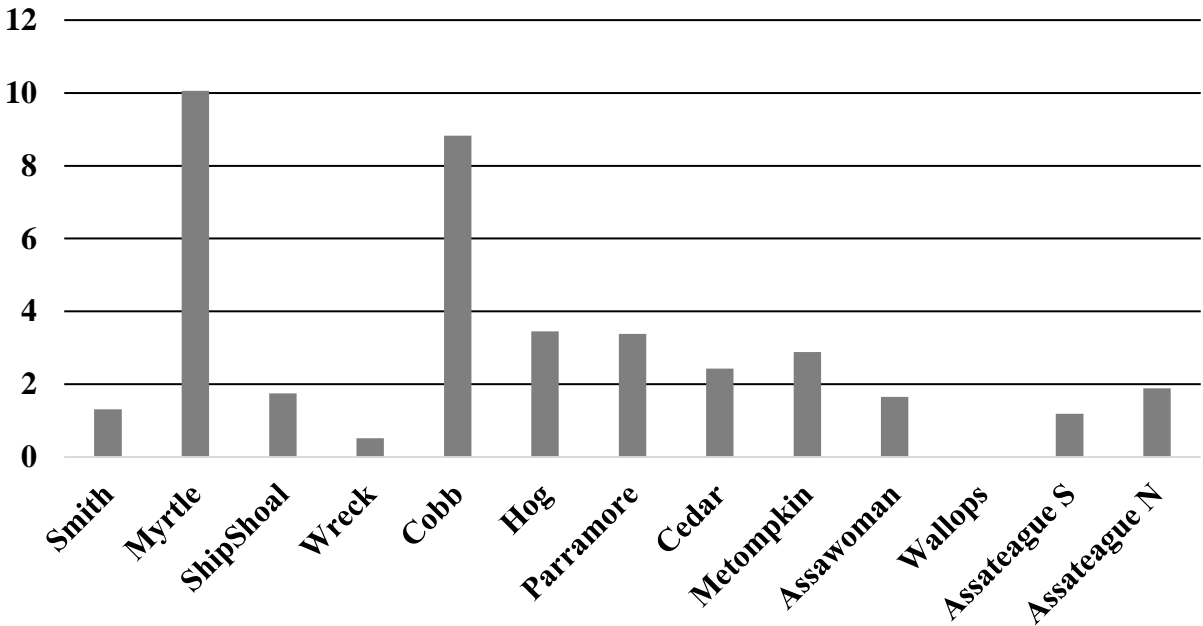

Accounting for different SLR rise scenarios and their contribution to SLRT in this region involves extracting the SLR and WE components from Equation 1. To achieve this we calculated a simple Pearson Correlation Coefficient that relates SLR to SLRT in this region from 1927 to present. We additionally utilized the tide information to depict major erosion events such as nor'easters and tropical storms as a proxy for WE. We then combined our storm measure with the SLR measure in a simple linear regression to predict SLRT. We find that the SLR combined with tropical storms accounts for between 0.4 and 0.8 of the variance in SLRT examined since 1927. We excluded from this analysis islands with constructed shoreline defenses where the relationship is close to 0. Knowing the approximate relationship between SLR and storms and SLRT allows us to vary each of these components to predict future shoreline retreat under the differing SLR scenarios in Table 1.

**DISCUSSION**

The static nature of the shorelines of the northern islands (Figure 2) should not be viewed as these islands not responding to SLR or being naturally resistant to SLR. The lack of SLRT is easily explained by the presence of substantial shoreline defenses along Wallops Island (King et al., 2010; Titus et al., 2010) likely built to protect critical NASA infrastructure on the island. The static nature of the shoreline is harder to explain on Assateague North, particularly as the Maryland portion of Assateague shows evidence of SLRT. One likely reason for the lack of substantial erosion along the majority of Assateague Island, VA is the regular beach nourishment that occurs north of the island in Ocean City, MD. It is possible that these nourished beaches replenish the shoreline of Assateague through longshore sediment supply. This explanation fits with the dominant longshore motion in the area, with sand from Ocean City possibly replenishing the beaches of Southern Assateague but bypassing the Maryland portions of Assateague. Aside from these northern islands, SLRT exists across all other islands.

Due to the highly-dynamic nature of barrier islands compared to other coastal lands and the high rates of uncertainty in the past, present, and future datasets no single representation of future shorelines will representatively depict their actual future location. In this paper, we have utilized data from 1850 to present to monitor the changes in the shorelines of VES barrier islands. The data reveals a pattern of shoreline retreat that is increasing towards present with increasing levels of SLRT likely post-2010. Differing islands, and even different transects, represent different levels of SLRT at you move along coastline and even across individual islands. The data presented though does show a pattern of consistent shoreline retreat at the transect, island, and study region scales and across all temporal domains allowing for an estimate 2100 CE shoreline to be developed. The 2100 shoreline presented contains a best-fit scenario taken from all transects, all islands, all temporal domains, all SLR scenarios and numerous subjective decisions.